\documentclass[pre, aps, amssymb, amsmath, nofootinbib, 10pt, twocolumn]{revtex4}
	
	\usepackage{graphicx}
	\usepackage{color}
	\usepackage{verbatim}
	\usepackage{subfigure}
	\include{subcaption}
	\usepackage{amssymb}

\begin{document}

	\title{Unification and new extensions of the no-pumping theorems of stochastic pumps }
	\author{Dibyendu Mandal}
	\affiliation{ 
	Helen Wills Neuroscience Institute, University of California, Berkeley, CA 94720, U.S.A.\\
	}

	\begin{abstract}	
	From molecular machines to quantum dots, a wide range of mesoscopic systems can be modeled by periodically driven Markov processes, or {\it stochastic pumps}.
	Currents in the stochastic pumps are delimited by an exact no-go condition called the no-pumping theorem (NPT).
	The letter presents a unified treatment of all the adaptations of NPT known so far, and further extends it to systems with many species of interacting particles.		
	\end{abstract}

	\maketitle

From the cargo transport to muscle contraction, a multitude of tasks inside our cells are performed by bimolecular complexes called the molecular motors. 
	They have inspired researchers to design artificial molecular complexes capable of controlled directed motion, such as translation along an axle~\cite{Panman2010} or a DNA origami track~\cite{Lund2010} and rotation along a molecular ring~\cite{Leigh2003}, among others~\cite{Kay2007, Bath2007, Feringa2007, Michl2009, Gu2010, Tierney2011}.
	Control technique of the artificial molecular machines is fundamentally different from  macroscopic machines, because the effects of inertia are negligible and friction and fluctuations play the dominant role in the molecular scale~\cite{Kay2007}. 
	An effective strategy is to periodically modulate the environment of the system to drive the system out of equilibrium and utilize the relaxation dynamics to generate the desired directed motion. 
	An example is provided by the experiments by Leigh {\it et al.}~\cite{Leigh2003} where an average directed rotation was induced in the artificial molecular machine [3]catenane by periodic modulation of temperature, radiation and chemical concentrations.

	Periodic modulation of external parameters to pump a desired directed current in stochastic systems is called stochastic pumping. 
	Success of this strategy is delimited by an exact condition, called the no-pumping theorem (NPT), which states that both the energy levels and the barriers of the system have to be varied to generate any directed current. 
	Consider, for example, the [2]catenane complex composed of a small molecular ring interlocked with a larger molecular ring, as depicted schematically in Fig.~1(a). 
	The numbers 1, 2, and 3 denote the metastable states of the smaller ring. 
	In Ref.~\cite{Leigh2003} the authors noted that the smaller ring could not be rotated along the larger ring unidirectionally just by the variation of the energies of the metastable states, even with the intuitively appealing strategy depicted in Fig.~1(b).
	 Such unexpected observations have led to a number of recent studies on stochastic pumps~\cite{Astumian2003, Astumian2007, Sinitsyn2007a, Sinitsyn2007b,  Rahav2008, Chernyak2008, Ohkubo2008, Chernyak2009, Sinitsyn2009, Horowitz2009, Maes2010, Astumian2011, Sinitsyn2011, Ren2011, Chernyak2011, Mandal2011, Mandal2012, Chernyak2012a, Chernyak2012b, Asban2014}. 
	
	\begin{figure}
	\label{fig:new}
	\includegraphics[trim = 0in 1in 1in 0.5in, width = 0.85 \columnwidth]{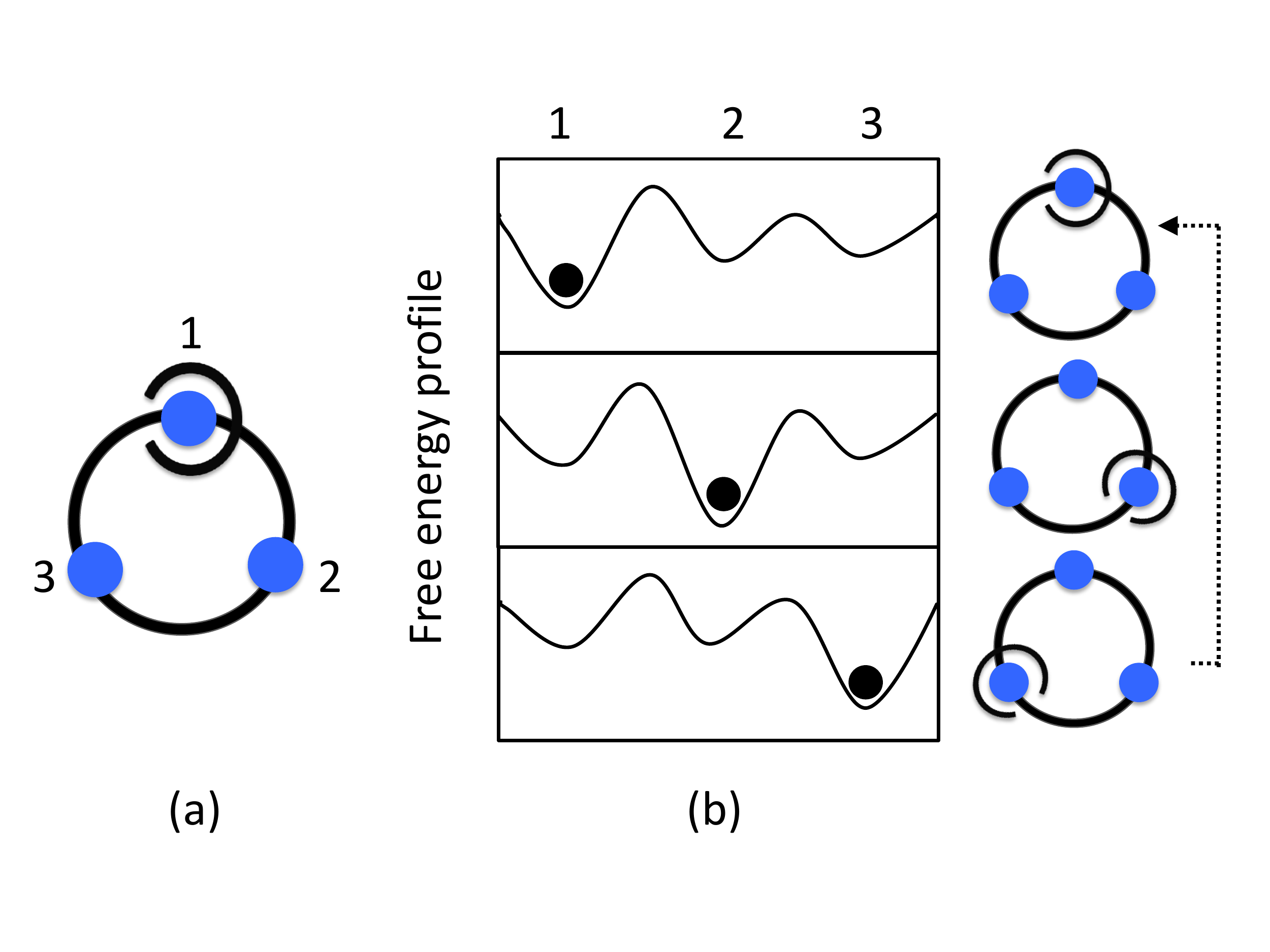}
	\caption{{\bf Illustration of NPT.} (a) Metastable states 1, 2, and 3 of the smaller ring. (b) Consider the strategy of periodic lowering the (free) energies of the metastable states in the clockwise sequence 1 $\rightarrow$ 2 $\rightarrow$ 3 $\rightarrow$ 1, as shown above. Intuitively, it is expected that the smaller ring will develop an average clockwise rotation as it follows the minimum energy state in the same sequence. Yet, according to NPT, there cannot be any such directed rotation if the barriers (local maxima) are kept fixed in time.}
	\end{figure}

	The NPT has been adapted to a number of scenario depending on the nature of the system.
	Starting with the original formulation for closed single-particle systems~\cite{Rahav2008}, it has been generalized to both closed and open many-particle systems~\cite{Chernyak2011, Asban2014}. 
	A curious feature of the open system NPT is that the particle reservoirs attached to the system need not be in equilibrium with each other at every moment; stochastic pumping will fail if just the time-averaged activities are the same (Eq.~\ref{eq:ActivityCond} below), along with other conditions of NPT~\cite{Chernyak2011}. 
	This is in contrast to the usual steady-state processes where the reservoirs at different chemical potentials lead to non-zero particle currents.

	The single-particle NPT follows from the closed-system many-particle NPT by assuming a priori that the system consists of a single particle. 
	The connection between the closed-system NPT and the open-system NPT is not so apparent, because the closed-system NPT does not follow from the open-system NPT in the limit of vanishing system-reservoir coupling, expected intuitively. 
	Further, the derivations of the two NPTs follow completely different routes: closed-system NPT follows from the evolution of the {\it state}-probabilities, dictated by appropriate master equations~\cite{Rahav2008, Asban2014}, whereas open-system NPT follows the {\it path}-probabilities of single-particle trajectories over large intervals of time~\cite{Chernyak2011}. 
	The letter uncovers the connection between the two results by giving a single mathematical framework that encompasses all the adaptations of NPT known so far. 
	Utilizing this framework, the letter also presents the appropriate generalization of NPT to stochastic pumps with many species of interacting particles. 
	Note that the following work focuses on the discrete-state description of stochastic pumps, as opposed to continuous-state descriptions~\cite{Reimann2002, Hanggi2009}.

	\begin{figure}
	\subfigure[]{
	\label{fig:setupa}
	\includegraphics[width = 0.45 \columnwidth]{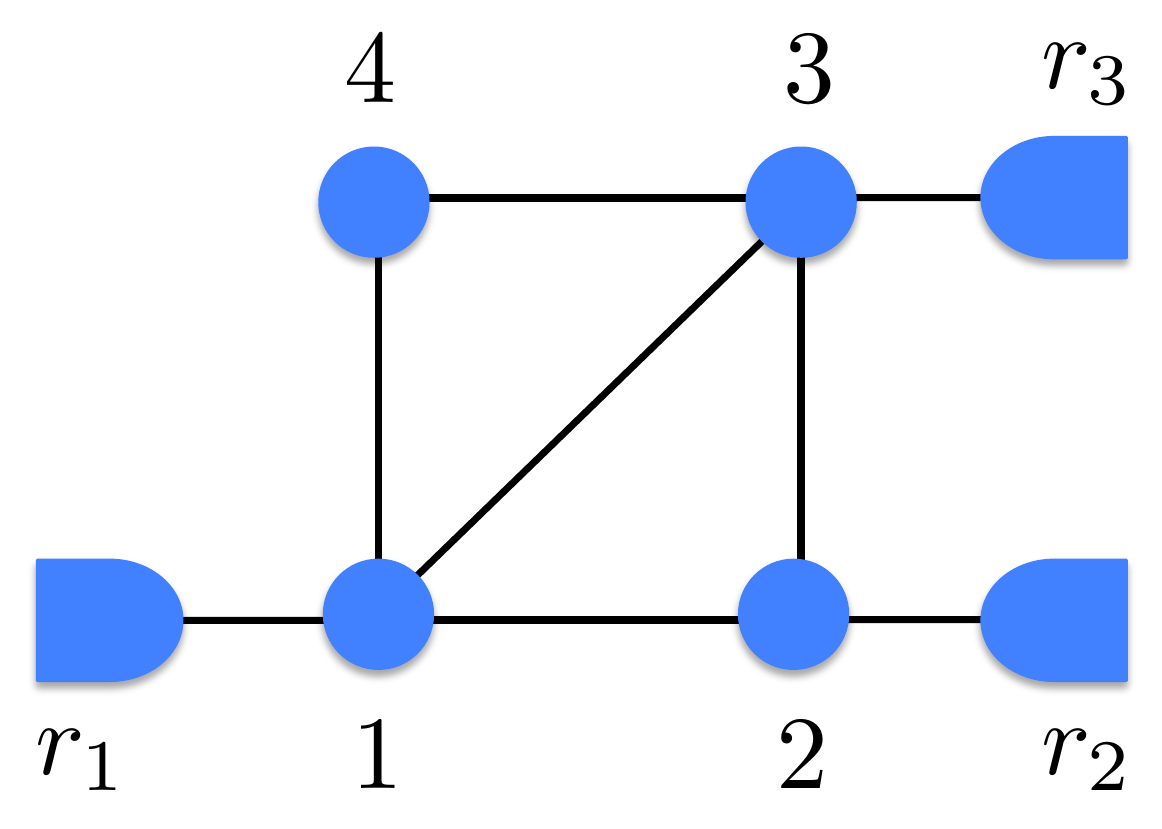}}
	\subfigure[]{
	\label{fig:setupb}
	\includegraphics[width = 0.45 \columnwidth]{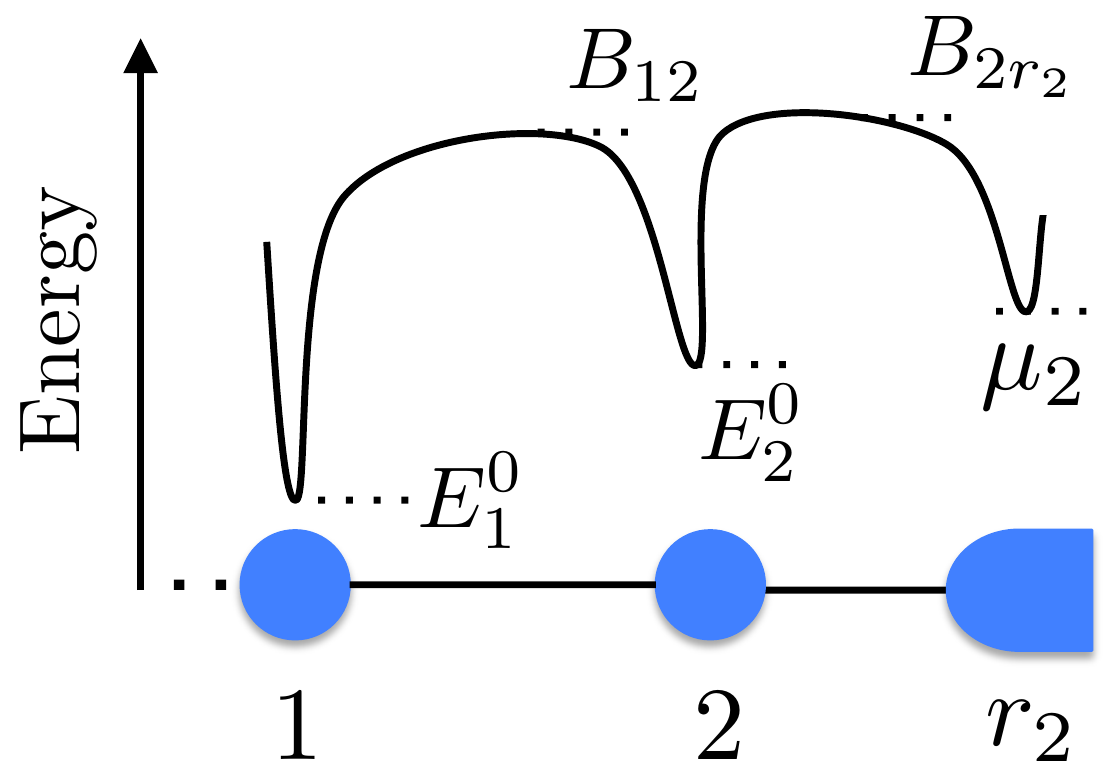}}
	\caption{{\bf Structure and energetics of an open system.} (a) An open system with 4 sites and 3 reservoirs. (b) Illustration of single-particle energies $\{E_i^0\}$, barriers $\{B_{ij}, B_{ir_i}\}$, and chemical potentials $\mu_i$.}
	\end{figure}

	\section{Model}

	Consider an open system consisting of $M$ physical sites, $i \in \{1, \ldots M\}$, all in contact with a thermal reservoir of absolute temperature $T$ and some in contact with respective particle reservoirs $r_i$ of chemical potentials $\mu_i$ [Fig.~\ref{fig:setupa}].
	For example, in quantum dot circuits in the Coulomb-blockade regime the dots act as the sites and the electrodes as the particle reservoirs~\cite{Bagrets2003, Jordan2004, Sukhorukov2007, Chernyak2011, Ganeshan2011}.  
	Driven by thermal fluctuations, particles make random transitions either between the system and the reservoirs or from one site to another. 
	These transitions are often modeled by Markov processes, each allowed transition being characterized by a positive (conditional) transition rate~\cite{Schnakenberg1976}.
	In the following, these rates are assumed to be both {\it ergodic} (any two sites are connected either directly or via intermediate sites) and {\it reversible} (any allowed transition implies the existence of its time-reversed transition) so that the system relaxes to a unique steady state from all initial conditions~\cite{vanKampen2007}.
	I also assume classical statistics and thermal units (Boltzmann constant equal to unity).

	The state of the system at any time $t$ is specified by the vector ${\bf n}(t) = [n_1(t), \ldots n_M(t)]$ where $n_i(t)$ denotes the instantaneous particle number in site $i$.
	At the ensemble level, the system is described by the instantaneous probability distribution $\{P_{\bf n}(t)\}$ which evolves according to the master equation~\cite{vanKampen2007}
	\begin{eqnarray}
	\label{eq:master1}
	\frac{d}{dt} P_{\bf n}(t) & = & \sum_{{\bf n'} \neq {\bf n}} J_{\bf n n'}(t), \\
	\label{eq:master2}
	J_{\bf n n'}(t) & = & R_{\bf n n'} P_{\bf n'}(t) - R_{\bf n' n} P_{\bf n}(t),
	\end{eqnarray}
where $J_{\bf n n'}$ is the probability current from state ${\bf n'}$ to ${\bf n}$ and $R_{\bf n n'}$ is the associated transition rate.

	From a thermodynamic point of view, the sites of the system correspond to the local (free) energy minima of the particles; see Fig.~2(b). 
	The master equation description (Eq.~\ref{eq:master1}) applies when the minima are deep enough with respect to the thermal fluctuations so that the particles inside the system almost always remain in the sites and the transitions among the sites (over the barriers) are instantaneous.
	Under these conditions only one particle may make a transition at any instant. 
	In other words, the rates $R_{\bf n n'}$ are nonzero only when ${\bf n}$ and ${\bf n'}$ differ by the placement of a single particle.
	If ${\bf n}$ is obtained from ${\bf n'}$ by placing a particle from site $j$ to site $i$, i.e., ${\bf n} = {\bf n'} + \hat{{\bf e}}_i - \hat{\bf e}_j$ where $(-) \hat{\bf e}_i$ denotes the entry (exit) of a particle into (from) site $i$, then $J_{\bf n n'}$ is a probability-current from $j$ to $i$.
	Similarly, for ${\bf n} = {\bf n'} \pm \hat{\bf e}_i$, $J_{\bf n n'}$ is a probability current from (to) reservoir $r_i$ to (from) site $i$.
	The average particle-currents, $J_{ij}(t)$ from site $j$ to site $i$, $J_{r_ii}$ from site $i$ to reservoir $r_i$, and $J_{ir_i}(t)$ from reservoir $r_i$ to site $i$, are then given by 
	\begin{eqnarray}
	\label{eq:Jij}
	J_{ij}(t) & = & \sum_{\bf n} J_{\bf n n'}(t), \quad {\bf n} = {\bf n'} +\hat{{\bf e}}_i -\hat{{\bf e}}_j, \\
	\label{eq:Jrii}
	J_{r_ii}(t) & = & \sum_{\bf n} J_{\bf n n'}(t), \quad {\bf n} = {\bf n'} -\hat{{\bf e}}_i , \\
	\label{eq:Jiri}
	J_{ir_i}(t) & = & \sum_{\bf n} J_{\bf n n'}(t), \quad {\bf n} = {\bf n'} +\hat{{\bf e}}_i . 
	\end{eqnarray}

	\subsection{Noninteracting case} 

	Consider the case where the particles do not interact with each other. 
	The energy of a particle in any site $i$ is denoted by $E^0_i$ and the barrier for its transitions from any site $i$ to another site $j$ is denoted by $B_{ij}$.
	If we assume that the system satisfies the conditions of detailed balance, that the current $J_{ij}$ are all zero in the steady state, then these barriers can be shown to be symmetric,  $B_{ij} = B_{ji}$~\cite{Mandal2011}. 
	The barrier between a reservoir $r_i$ and the associated site $i$ is similarly denoted by $B_{ir_i} = B_{r_i i}$. 
	These single-particle energy parameters are illustrated in Fig.~\ref{fig:setupb}. 
	In terms of these parameters, the many-particle transitions rates $R_{\bf n n'}$ are given by the following Arrhenius forms,
	\begin{eqnarray}
	\label{eq:Arr01}
	R^0_{{\bf n} + \hat{{\bf e}}_i -\hat{{\bf e}}_j, {\bf n}} & = & \nu \, n_j \exp{-[\beta (B_{ij} - E_j^0})], \\
	\label{eq:Arr02}
	R^0_{{\bf n} - \hat{{\bf e}}_i, {\bf n}} & = & \nu \, n_i \exp{-[\beta (B_{ir_i} - E_i^0)]}, \\
	\label{eq:Arr03}
	R^0_{{\bf n} +\hat{{\bf e}}_i, {\bf n}} & = & \nu \, \exp{-[\beta (B_{ir_i} - \mu_i)]}, 
	\end{eqnarray} 
where $\nu$ is a frequency factor and $\beta = 1/T$ is the inverse temperature. 
	The factors $n_j$ and $n_i$, in Eqs.~\ref{eq:Arr01} and \ref{eq:Arr02}, respectively, indicate the generic fact that, in any state ${\bf n}$,  $n_k$ particles contribute to currents out of site $k$.

	\subsection{Interacting case} 

	Consider now a short-range interaction among the particles, so that the particles in the same site can interact with one another. 
	In terms of the thermodynamic picture introduced above, the range of the interactions is assumed to be much smaller than the separations among the energy minima. 
	The interactions are also assumed to be sufficiently weak so that the stable configurations of the non-interacting case do not change, i.e., states are neither created nor annihilated by the interactions.
	Then, the total energy at any site $i$ is given by 
	\begin{equation}
	\label{eq:site_energy}
	E_i(n_i) = E^0_i n_i + U_{i}^\text{ int}(n_i),
	\end{equation}
where $U_i^{\text{ int}}(n_i)$ is the site-dependent interaction energy of $n_i$ particles. 
	Because of the short range of the interactions, a particle during a transition (crossing a barrier) does not interact with the other particles, which are in their respective sites.
	Consequently, the barriers $\{B_{ij}, B_{ir_i}\}$ are unaffected by the interactions, a standard assumption in the many-particle NPTs and zero-range processes~\cite{Evans2005}.

	 To explore the effects of interactions on the transition rates $R_{\bf n n'}$, for any state ${\bf n}$ consider a configuration of particles ${\bf X} = (x_1, \ldots x_{\sum n_i})$ where $x_k$ denotes the site of the $k$-th particle. 
	A transition from ${\bf n}$ to (${\bf n}+\hat{{\bf e}}_i-\hat{{\bf e}}_j$) corresponds a particle in $x_k = j$ jumping to site $i$. 
	If ${\bf X'}$ is a new configuration, having $n_j$ possible values corresponding to $n_j$ possible particles that could have made the transition, the transition rate $R_{\bf X', X}$ is given by the Arrhenius form
	\begin{equation}
	\label{eq:ArrMicro}
	R_{\bf X', X} = \nu \exp{-[\beta( B_{\bf X', X} - E_{\bf X})]},
	\end{equation}
where $B_{\bf X', X} = B_{\bf X, X'}$ is the barrier between configurations ${\bf X}$ and ${\bf X'}$ and $E_{\bf X}$ is the energy of the configuration ${\bf X}$ (or equivalently of state ${\bf n}$). 
	The barrier $B_{\bf X', X}$ is the instantaneous energy of the system when a particle with $x_k = j$ moves on to the single-particle barrier $B_{ij}$, on its way to $i$, and the other particles remain in their sites.
	Accordingly, we have 
	\begin{eqnarray}
	\label{eq:BMany}
	B_{\bf X', X} & = & B_{ij} + \sum_{k \neq j} E_k(n_k) + E_j(n_j - 1), \\
	\label{eq:EMany}
	E_{\bf X} & = & \sum_k E_k(n_k).
	\end{eqnarray}
	Because there are $n_j$ possible values for the configuration ${\bf X'}$, the transition rate from ${\bf n}$ to (${\bf n} +\hat{{\bf e}}_i- \hat{{\bf e}}_j$) is given by $n_j$ times $R_{\bf X', X} $.
	Then, Eqs.~\ref{eq:site_energy}--\ref{eq:EMany} lead to, after some algebra,
	\begin{equation} 
	\label{eq:Arr11}
	R_{{\bf n} + \hat{{\bf e}}_i - \hat{{\bf e}}_j, {\bf n}} = e^{-\beta B_{ij}} f_j(n_j)
	\end{equation}	
with  $f_j (n_j)  =  n_j \, \nu \exp{\beta \left[E^0_{j} + U_j^\text{int}(n_j) - U_j^\text{int}(n_j - 1)\right]}$.
	Similarly, the expressions for the other rates can be obtained as
	\begin{eqnarray}
	\label{eq:Arr12}
	R_{{\bf n}-\hat{{\bf e}}_i, {\bf n}} & = & e^{- \beta B_{ir_i}} f_i(n_i),\\ 
	\label{eq:Arr13}
	R_{{\bf n}+\hat{{\bf e}}_i, {\bf n}} & = & \nu \, \exp{-[\beta (B_{ir_i} - \mu_i)]} =  R^0_{{\bf n}+\hat{{\bf e}}_i, {\bf n}}.
	\end{eqnarray}

	Combining Eqs.~\ref{eq:master2} -- \ref{eq:Jiri}, \ref{eq:Arr11} -- \ref{eq:Arr13} and using the symmetry of the barriers $\{B_{ij} = B_{ji}, B_{ir_i} = B_{r_ii}\}$, we can derive the following expressions for the currents (omitting the factors of $\beta$ and $\nu$ henceforth for clarity)
	\begin{eqnarray}
	\label{eq:JijDB}
J_{ij}(t) & =  & \sum_{\bf n} P_{\bf n}(t) \, e^{-B_{ij}} \left[ f_j(n_j) - f_i(n_i)\right], \\ 
	\label{eq:JriiDB}
J_{r_ii}(t) & = & \sum_{\bf n} P_{\bf n}(t) \, e^{-B_{ir_i}} \left[f_i(n_i)  - e^{\mu_i}\right], \\
	\label{eq:JiriDB}
J_{i r_i}(t) & = & \sum_{\bf n} P_{\bf n}(t) \, e^{-B_{ir_i}} \left[ e^{\mu_i} - f_i(n_i) \right].
	\end{eqnarray}

	Stochastic pumping concerns periodic variation of the parameters $\{E_i^0$, $U_i^\text{int}$, $B_{ij}$, $B_i$, $\mu_i$, $\nu\}$, with some period $\tau$, and consequent transport of particles. 
	The system relaxes to a unique periodic steady state, $P^\text{ps}_{\bf n}(t+\tau) = P^\text{ps}_{\bf n}(t)$ for all ${\bf n}$~\cite{Jung1993}, to be denoted with a superscript ``ps."
	The total transport of particles over a time-period $\tau$ is given by the integrals 
	\begin{equation}
	\label{eq:Phi}
	\Phi_{ij}^\text{ps} = \int_\tau J_{ij}^\text{ps}(t), \quad \Phi_{r_ii}^\text{ps} = \int_\tau J_{r_ii}^\text{ps}(t), \quad \Phi_{ir_i}^\text{ps} = \int_\tau J_{ir_i}^\text{ps}(t).
	\end{equation} 
	For example, $\Phi_{12}^\text{ps}$ gives, on the average, the net number of particles hopping from site 2 to site 1 over a time-period $\tau$. 
	In the catenane experiment~\cite{Leigh2003}, the integrated currents correspond to the net number of rotations of the smaller rings around the larger ring.

	\section{NPT}
	\label{sec:NPT}

	According to NPT, not all pumping will lead to non-zero integrated currents, because all the integrated currents $\{\Phi_{ij}^{\text ps}, \Phi_{r_ii}^\text{ps}, \Phi_{ir_i}^{\text ps}\}$ are zero if either of the following generic conditions is met:
	
	(i) Only the barriers $\{B_{ij}(t), B_{ir_i}(t)\}$ are varied in time keeping all the site energies $\{E^0_i, U_i^{\text{int}}\}$ fixed and the chemical potentials fixed and uniform, $\mu_i = \mu$, same for all $r_i$.
	
	(ii) Only the site energies and chemical potentials $\{E^0_i(t)$, $U_i^{\text{int}}$, $\mu_i(t)\}$ are varied in time, keeping all the barriers $\{B_{ij}, B_i\}$ fixed and the {\it time-averaged activities} of the reservoirs uniform,
	\begin{equation}
	\label{eq:ActivityCond}
	\frac{1}{\tau} \int_\tau \mathrm{d}t \, \exp{\mu_i(t)} = \exp{\overline{\mu}}, \text{ same for all } r_i.
	\end{equation}

	Condition (i) is easy to understand. 
	In this case, the system relaxes to the Boltzmann distribution, $P^\text{eq}(n_i = m) \propto e^{ - E_i(m)}$, where all the instantaneous currents $\{J_{ij}^\text{ps}, J_{r_ii}^\text{ps}(t), J_{ir_i}^\text{ps}\}$, and therefore their integrals $\{\Phi_{ij}^\text{ps}, \Phi_{r_ii}^\text{ps}, \Phi_{ir_i}^\text{ps}\}$, are zero.
	In the following, therefore, only the condition (ii) is considered. 
	As noted before, condition (ii) does not require the chemical potentials to be uniform at all times, only the time-averaged activities have to be the same.  
	This is satisfied, for example, in the case of sinusoidal variation of the chemical potentials with equal amplitude but arbitrary phase differences. 
	Note also that the conditions of NPT are {\it sufficient}.
	There may be situations, possibly accidental, where the integrated currents are all zero even when neither of the two conditions is met.

	\section{Proof}
	\label{sec:Proof}

\begin{figure}[tbp]
	\subfigure[]{
	\label{fig:cyc}
	\includegraphics[width = 0.45 \columnwidth]{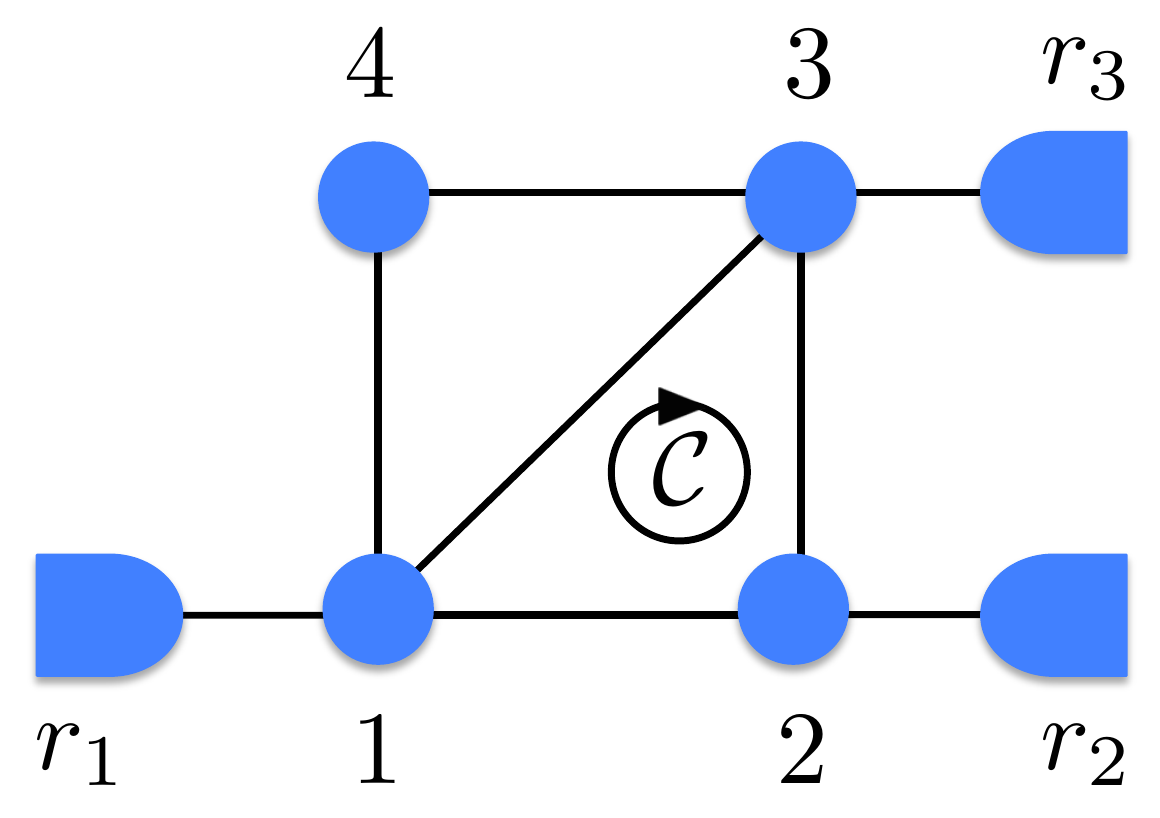}}
	\subfigure[]{
	\label{fig:path}
	\includegraphics[width = 0.45 \columnwidth]{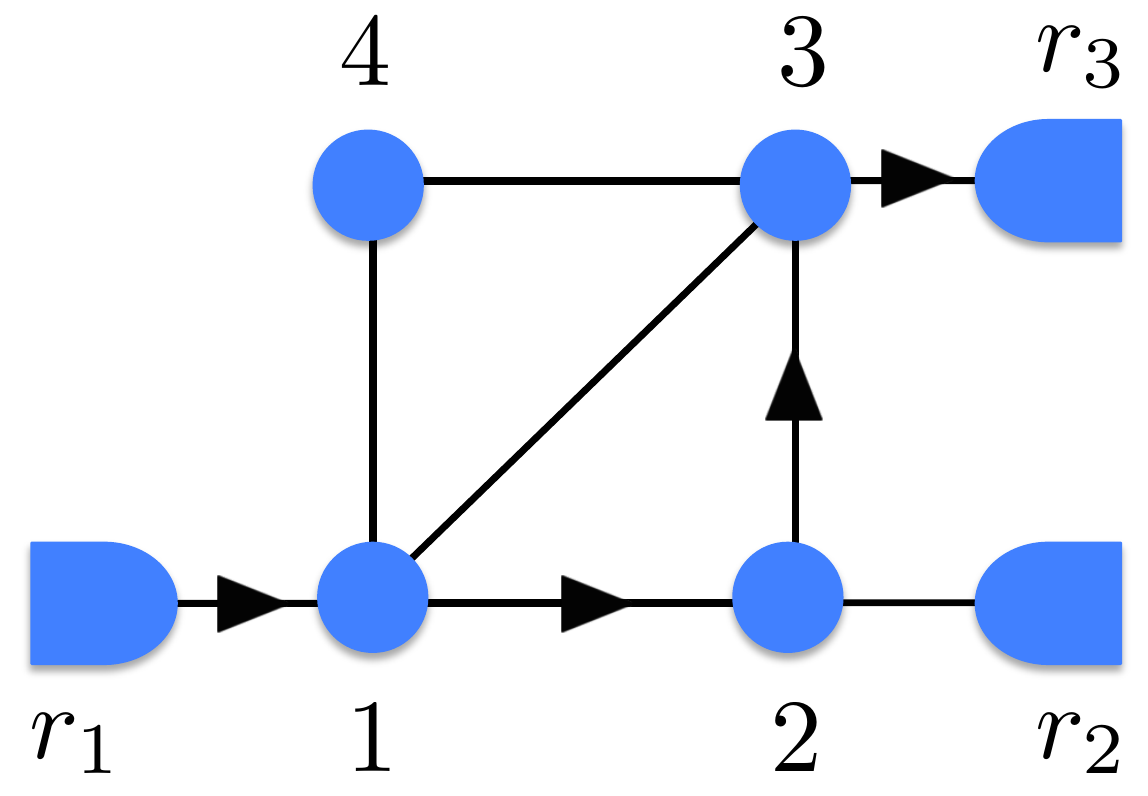}}
	\caption{{\bf Cycles and arrows.} (a) The sequence $\{1,3,2,1\}$ forms a cycle $\cal{C}$. (b) The arrows imply that the integrated currents along the corresponding edges, $\{\Phi^\text{ps}_{1r_1}, \Phi^\text{ps}_{21}, \Phi^\text{ps}_{32}, \Phi^\text{ps}_{r_33}\}$, are all positive.}
	\end{figure}

	The NPT is a consequence of two physical conditions: (1) periodicity of stochastic pumps, $P_{\bf n}^{\text{ps}}(t + \tau) = P_{\bf n}^{\text{ps}}(t)$ for all ${\bf n}$, and (2) the conditions of detailed balance, translating into the symmetry of the one-particle barriers $\{B_{ij} = B_{ji}, B_{ir_i} = B_{r_ii}\}$ (or equivalently, Eqs.~\ref{eq:JijDB} -- \ref{eq:JiriDB} for the currents).
	Because of the periodicity condition, the average number of particles at any site $i$ is a periodic function of time, so the total integrated current into any site $i$ must be zero,
	\begin{equation}
	\label{eq:Cont_Open}
\Phi_{i r_i}^\text{ps} + \sum_{j \neq i} \Phi^\text{ps}_{ij} = 0.
	\end{equation} 	
	Similarly, from the periodicity of the system as a whole, we have 				
	\begin{equation}
	\label{eq:ContOpenWhole}
\sum_{i = 1}^M \Phi_{i r_i}^\text{ps} = 0.
	\end{equation} 		
	Detailed balance conditions put restrictions on the currents over cycles and paths. 
	For a sequence of sites $\{i_1, \ldots, i_n, i_{n+1} \equiv i_1\}$ that form a cycle ${\mathcal C}$, as shown in Fig.~\ref{fig:cyc}, using Eq.~\ref{eq:JijDB} repeatedly for each of the consecutive pairs of sites, {\it in the same cyclic order}, we can obtain~\cite{Chernyak2009}
	\begin{equation}
	\label{eq:DBCyc}
\sum_{m = 1}^{n} \exp{(B_{i_m i_{m+1}})} \, J_{i_m i_{m+1}}(t)  = 0.
	\end{equation}   
	Because the barriers $\{B_{ij}\}$ are kept fixed in time, we can integrate Eq.~\ref{eq:DBCyc} over a time-period $\tau$ in the periodic steady state to obtain
	\begin{equation}
	\label{eq:DBCycPhi}
\sum_{m = 1}^{n} \exp{(B_{i_m i_{m+1}})} \, \Phi^\text{ps}_{i_m i_{m+1}} = 0.
	\end{equation}   
	For any path $\{r_i$, $i,$ $k,$ $\ldots,$ $l,$ $j,$ $r_j\}$ from any reservoir $r_i$ to another reservoir $r_j$ we can further obtain
	\begin{equation}
	\label{eq:DBPathPhi}
e^{B_{ir_i}} \Phi^\text{ps}_{ir_i}  + e^{B_{ik}} \Phi^\text{ps}_{ik} + \ldots + e^{B_{lj}} \Phi^\text{ps}_{lj} + e^{B_{r_jj}} \Phi^\text{ps}_{r_jj} = 0.
	\end{equation}
	First, using the expression of currents, Eqs.~\ref{eq:JijDB}--\ref{eq:JiriDB}, we get the linear relation: $e^{B_{ir_i}} J_{ir_i}  + e^{B_{ik}} J_{ik} + \ldots + e^{B_{lj}} J_{lj} + e^{B_{r_jj}} J_{r_jj} = e^{\mu_i(t)} - e^{\mu_j(t)}$. 
	Then, on integration over $\tau$ and from the condition of uniform time-averaged activities, Eq.~\ref{eq:ActivityCond}, we get Eq.~\ref{eq:DBPathPhi}.

	Consider the case where all the reservoir integrated currents $\{\Phi_{ir_i}, \Phi_{r_ii} \}$ happen to be zero, similar to that of a closed system.
	As the corresponding analysis has been presented before, within the framework of closed many-particle stochastic pumps~\cite{Asban2014}, it is omitted in the following.
	Consider now the case where at least one reservoir integrated current is nonzero, say $\Phi_{ir_i} > 0$.
	I have chosen the sign arbitrarily as it does not have any essential bearing on the proof. 
	Because of the periodicity condition~\ref{eq:ContOpenWhole}, there must exist at least another reservoir current with opposite sign, say $\Phi_{ir_j} < 0$. 
	In fact, there must exist a path from reservoir $r_i$ to some reservoir $r_j$ such that the integrated currents along the corresponding edges are all positive.
	Otherwise, the particles injected from $r_i$ will continually accumulate in the system, violating the periodicity conditions~\ref{eq:Cont_Open}.
	This intuition can be formalized utilizing the construction suggested in Refs.~\cite{Mandal2011, Mandal2012}. 
	First, one draws an arrow along the direction of each positive integrated current; see Fig.~\ref{fig:path} for an illustration. 
	Then, one constructs a set $\mathcal{D}_i$ comprising of all the sites and reservoirs that can be reached from the site $i$ following the arrows. 
	For example, in Fig.~\ref{fig:path} we have ${\mathcal D}_1 = \{2, 3, r_3\}$.
	The set $\mathcal{D}_i$ must contain at least one reservoir $r_j$ which can ``absorb" the particles injected from reservoir $r_i$; otherwise, the total number of particles in the sites of $\mathcal{D}_i$ will increase indefinitely in time violating the periodicity conditions~\ref{eq:Cont_Open}.
	We can now construct a path from $r_i$ to $r_j$, using the elements of $\mathcal{D}_i$, such that all the arrows along the path point in the same direction, i.e., the integrated currents along the path are all positive, by construction. 
	But this will contradict Eq.~\ref{eq:DBPathPhi}, according to which not all integrated currents along such a path can be positive.
	We must, therefore, conclude that there could not have been any nonzero reservoir integrated current at the first place, thus, completing the proof.

	\section{Unification}
	\label{sec:Uni}

	As noted above, the closed-system NPT forms an integral part of the open-system NPT. 
	If we assume, a priori, that the reservoir currents are all zero, Eqs.~\ref{eq:Cont_Open} and \ref{eq:DBCycPhi} survive and they suffice to prove the closed-system NPT~\cite{Asban2014}.
	This observation leads to a common conceptual origin for the closed- and the open-system NPTs: They both appear as consequences of the same pair of physical principles -- the periodicity of stochastic pumps and the conditions of detailed balance (translating into the symmetry of the single-particle barriers).

	\section{New extensions}
	\label{sec:New}

	Consider an open system with $M$ physical sites and many species of interacting particles, $\alpha = 1, \ldots N$, each having its own reservoirs and parameters $\{$$E_i^{0, \alpha}$, $\mu_i^\alpha$, $B_{ij}^\alpha$, etc.$\}$. 
	The interaction energy at any site $i$, $U_i^\text{int}({\bf n}_i)$, is a function of the site-composition ${\bf n}_i = (n_i^1, \ldots n_i^N)$, where $n_i^\alpha$ denotes the number of $\alpha$ particles in site $i$.
	The state of the whole system is given by the set $S = \{{\bf n}_1, \ldots {\bf n}_M\}$.
	For any species $\alpha$, the particle-currents $\{J_{ij}^\alpha, J_{r_ii}^\alpha, J_{ir_i}^\alpha\}$ are given by
	\begin{eqnarray}
	\label{eq:JijAlpha}
J_{ij}^\alpha(t) & = & \sum_S P_S(t) \, e^{-B_{ij}^\alpha} \left[ f_j^\alpha({\bf n}_j) - f_i^\alpha({\bf n}_i)   \right], \quad  \\ 
	\label{eq:JriiAlpha}
J_{r_i i}^\alpha(t) & = & \sum_S P_S(t) \, e^{- B^\alpha_{ir_i}} \left[f_i^\alpha({\bf n}_i) - e^{\mu^\alpha_i}\right], \\
	\label{eq:JiriAlpha}
J_{i r_i}^\alpha(t) & = & \sum_S P_S(t) \, e^{- B^\alpha_{ir_i}} \left[ e^{\mu^\alpha_i} - f_i^\alpha({\bf n}_i) \right],
	\end{eqnarray}
with $f_i^\alpha({\bf n}_i) = n^\alpha_i  \exp{\left[E_i^{0, \alpha} + U_i^{\text{int}}({\bf n}_i) -  U_i^{\text{int}}({\bf n}_i - \hat{{\bf e}}^\alpha_i)\right]}$, where $\hat{{\bf e}}^\alpha_i$ denotes the entry of an $\alpha$  particle into site $i$.
	Equations~\ref{eq:JijAlpha}--\ref{eq:JiriAlpha} are the generalizations of Eqs.~\ref{eq:JijDB}--\ref{eq:JiriDB}, respectively, to the multi-species scenario.
	In case of stochastic pumping, we have the species-specific integrated currents 
	\begin{equation}
	\Phi_{ij}^{\alpha, \text{ps}} = \int_\tau J_{ij}^{\alpha,\text{ps}}, \quad \Phi_{r_ii}^{\alpha, \text{ps}} = \int_\tau J_{r_i i}^{\alpha, \text{ps}}, \quad \Phi_{ir_i}^{\alpha, \text{ps}} = \int_\tau J_{ir_i}^{\alpha, \text{ps}}.
	\end{equation}

	It is now possible to formulate an NPT for each species $\alpha$: All the $\alpha$-particle integrated currents, $\{\Phi_{ij}^{\alpha, \text{ps}}, \Phi_{r_ii}^{\alpha, \text{ps}}, \Phi_{ir_i}^{\alpha, \text{ps}}\}$, are zero if either (i) all the energy parameters $\{E_i^{0, \alpha}, U_i^\text{int}, \mu_i^\alpha\}$ of the $\alpha$-species are fixed in time and the corresponding chemical potentials uniform, $\mu_i^\alpha = \mu^\alpha$ for all $r^\alpha_i$, or (ii) all its the barriers $\{B_{ij}^\alpha, B_{ir_i}^\alpha\}$ are fixed in time and the average activities uniform $(1 / \tau ) \int_\tau {\rm d}t \, e^{ \mu_i^\alpha (t)} = e^{\overline{\mu^\alpha}}$, same for all $r^\alpha_i$, {\it irrespective of the other parameters}.

	The existence of such species-specific NPTs in multi-species stochastic pumps can be inferred from the following steps of argument.
	[I consider only the nontrivial condition (ii).]
	In the periodic steady state the average number of particles of any species $\alpha$ in any site $i$ is a periodic function of time.
	For each $\alpha$, therefore, we have the following periodicity conditions
	\begin{equation}	
	\sum_{j \neq i} \Phi_{ij}^{\alpha, \text{ps}} + \Phi_{ir_i}^{\alpha, \text{ps}} = 0, \quad \sum_{i = 1}^M \Phi_{ir_i}^{\alpha, \text{ps}} = 0,
	\end{equation} 
just like Eqs.~\ref{eq:Cont_Open} and \ref{eq:ContOpenWhole} before.
	From the formal resemblance of the currents in Eqs.~\ref{eq:JijAlpha}--\ref{eq:JiriAlpha} with those of Eqs.~\ref{eq:JijDB}--\ref{eq:JiriDB}, respectively, one can also derive the analogues of Eqs.~\ref{eq:DBCycPhi} and \ref{eq:DBPathPhi} for each $\alpha$: 
	\begin{equation}
	\sum_{m = 1}^{n} \exp{(B^\alpha_{i_m i_{m+1}})} \, \Phi^{\alpha, \text{ps}}_{i_m i_{m+1}} = 0 \quad \text{and}
	\end{equation}
	\begin{equation} 
	e^{B^\alpha_{ir_i}} \Phi^{\alpha, \text{ps}}_{ir_i} + e^{B^\alpha_{ik}} \Phi^{\alpha, \text{ps}}_{ik} + \ldots + e^{B^\alpha_{lj}} \Phi^{\alpha, \text{ps}}_{lj} + e^{B^\alpha_{r_jj}} \Phi^{\alpha, \text{ps}}_{r_jj} = 0,
	\end{equation}
for each cycle ${\cal C} = \{ i_1, \ldots, i_n, i_{n+1} \equiv i_1\}$ and  for each path from any reservoir $r_i^\alpha$ to another reservoir $r_j^\alpha$, respectively.
	The $\alpha$-particle NPT then follows exactly the same way as in the above proof.
	By corollary, an NPT also holds for the sum of the integrated currents of all the species that individually satisfy the conditions of NPT.

	As before, the closed-system species-specific NPT can be obtained from the above analyses by assuming, a priori, that the reservoir currents $\{\Phi_{ir_i}^{\alpha, \text{ps}}, \Phi_{r_ii}^{\alpha, \text{ps}}\}$ are all zero.

	\begin{figure}[tbp]
	\includegraphics[width = .9 \columnwidth]{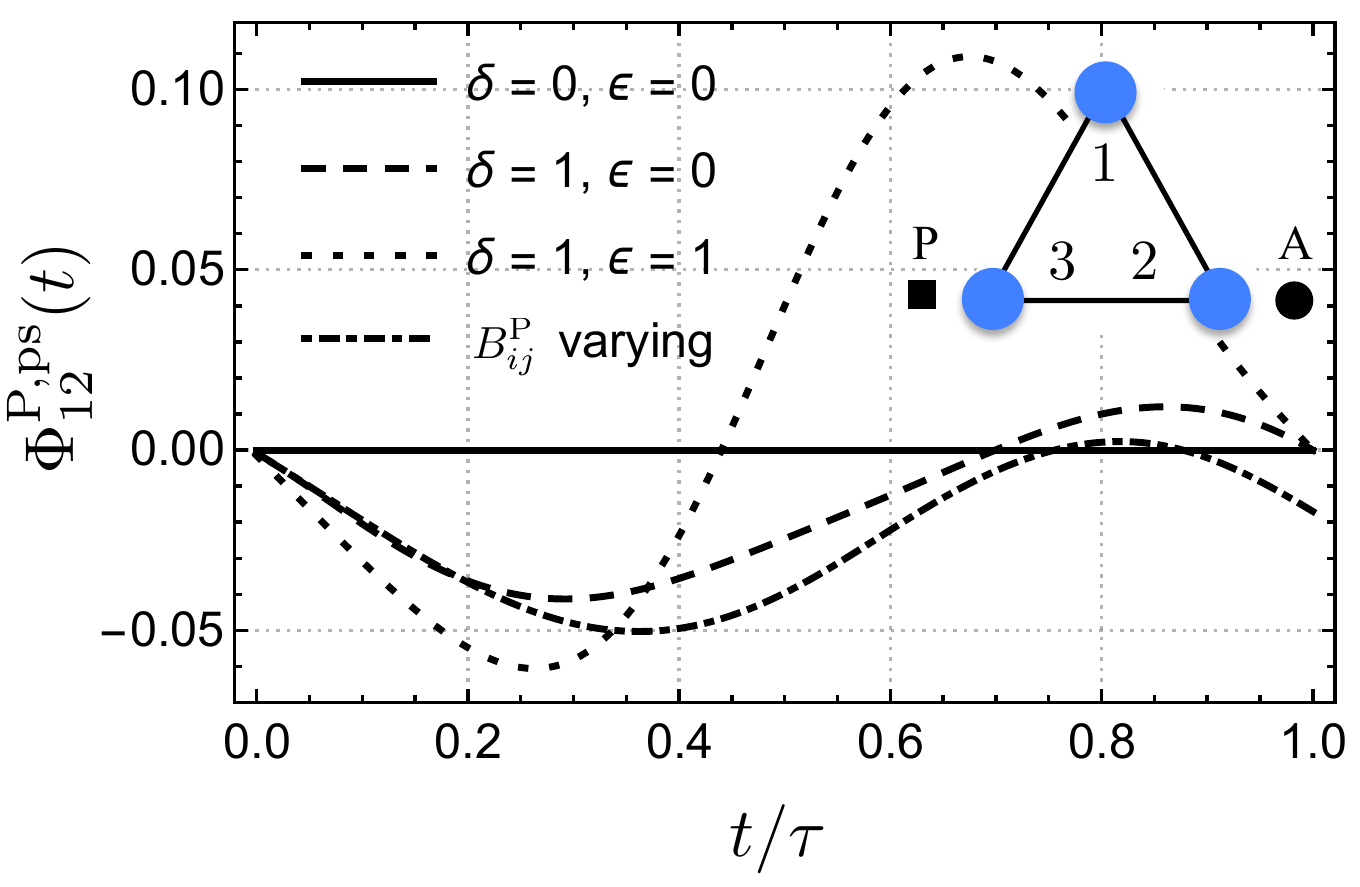}
	\caption{{\bf An illustration of the species-specific NPTs.} Interactions alone cannot drive the passive (P) particle along the cycle; its barrier(s) need to be varied. (See text for details.)}
	\label{fig:example}
	\end{figure}

	\section{Illustration} 

	Consider the case where one tries to control the dynamics of a relatively passive species of particles, with less control on their parameters, via an active species (inspired by studies in~\cite{Savelev2003, Savelev2004}).	
	For simplicity, let us assume the system to be closed, and consisting of just three sites  $\{1, 2, 3\}$ and two particles A (active) and P (passive), as depicted in the top right corner of Fig.~\ref{fig:example}. 
	Let us consider the following parametric values: $\tau = 10$; $\nu^\text{A}$, $\nu^\text{P}$, $T = 1$; $ E_i^{0, \text{A}}(t)$ $=$ $ - 2 + \cos{\left[2 \pi \left( \frac{t}{\tau} + \frac{i - 1}{3}\right) \right]}$; $B^\text{A}_{i, i+1}(t)= 2 + E^{0,\text{A}}_{i+1}(t)$ with $i + 1 = 1 \text{ for } i = 3)$; $\{E^{0, \text{P}}_1, E^{0, \text{P}}_2, E^{0, \text{P}}_3 \}$ $=$ $\{-0.1, -0.3, -0.2\}$; and $B^\text{P}_{ij} = 0$ for all $i, j$. 
	Note that both the energies and the barriers of the active particle are varied in time (leading to a non-zero integrated current $\Phi_{i, i+1}^\text{A,ps} = -0.12$), while all the parameters of the passive particle are fixed because they are assumed to be ``inaccessible." 
	Consequently, the passive particle integrated currents are zero (solid line in Fig.~\ref{fig:example}). 
	Intuitively, it is now expected that the active particle will be able to induce a directed current in the passive particle if there is a non-zero interaction between them: For repulsive interaction, the active particle will ``push" the latter forward, and for attractive interaction it will ``pull" the other along. 
	However, this induced transport is forbidden by the new NPT: Because all the barriers of the passive particle are fixed in time, condition (ii) applies and there can be no integrated current, even when the interaction between the two particles is time-dependent. 
	To see this consider the following form of interaction: $U_i^\text{int}$ $=$ $-\delta + \epsilon \cos{\left[2 \pi \left( \frac{t}{\tau} + \frac{i - 1}{3}\right) \right]}$. 
	We have indicated the results in Fig.~\ref{fig:example}, plotting the integrated current of the passive particle from site 2 to 1 up to  different portions $t$ of a time-period $\tau$: $\Phi_{12}^\text{P,ps}(t)$ $=$ $\int_0^t {\rm d} s\, J_{12}^\text{P, ps}(s)$ for $0 \leq t \leq \tau$. 
	 The dashed and the dotted lines correspond to the time-independent ($\epsilon = 0$) and the time-dependent ($\epsilon \neq 0$) interactions, respectively. 
	 In both cases, the passive particle sloshes back and forth during the time-period, because of its interaction with the active particle, but at the end of the time-period the net integrated current is still zero. 
	 Only when the barriers of the passive particle are varied in time in addition to the interaction ($\epsilon \neq 0$), thus violating the conditions of NPT, that we get a non-zero integrated current in the end (the dot-dashed curve).

	By unifying all previous results on NPT into a single theoretical framework the current work uncovers their essential conceptual unity. 
	Extension of NPT to interacting many species of particles further illustrates the generality of this result, and will prove useful when, for example, many types of artificial molecular machines are involved in a single setup.
	Finally, the result emphasizes the role of the barriers in stochastic pumps: While they  do not affect the equilibrium distributions, the barriers do play a critical role in the out-of-equilibrium currents.

	 \acknowledgments
	I gratefully acknowledge many useful discussions with Christopher Jarzynski, Royce K. P. Zia, Nikolai Sinitsyn, and Zhiyue Lu, and financial support from the National Science Foundation (USA) under the grant OCE 1245944.

	\end{document}